# Kinematics of the diffuse intragroup and intracluster light in groups and clusters of galaxies in the Local Universe within 100 Mpc distance

**Magda Arnaboldi**[1*], **Ortwin E. Gerhard**[2]

[1]European Southern Observatory (ESO), Garching, Germany

[2]Max-Planck-Institute für extraterrestrische Physik, Garching, Germany

**\* Correspondence:**
Corresponding Author: marnabol@eso.org

**Keywords: galaxies: halos, evolution. Galaxies: clusters: diffuse component, interactions. planetary nebulae: general**

## Abstract

Nearly all intragroup (IGL) and intracluster light (ICL) comes from stars that are not bound to any single galaxy but were formed in galaxies and later unbound from them. In this review we focus on the physical properties - phase space properties, metallicity and age distribution - of the ICL and IGL components of the groups and clusters in the local universe, within 100 Mpc distance. Kinematic - information on these very low surface brightness structures mostly comes from discrete tracers such as planetary nebulae and globular clusters, showing highly unrelaxed velocity distributions. Cosmological hydrodynamical simulations provide key predictions for the dynamical state of IGL and ICL and find that most IC stars are dissolved from galaxies that subsequently merge with the central galaxy. The increase of the measured velocity dispersion with radius in the outer halos of bright galaxies is a physical feature that makes it possible to identify IGL and ICL components. In the local groups and clusters, IGL and ICL are located in the dense regions of these structures. Their light fractions relative to the total luminosity of the satellite galaxies in a given group or cluster are between a few to ten percent, significantly lower than the average values in more evolved, more distant clusters. IGL and ICL in the Leo I and M49 groups, and the Virgo cluster core around M87, has been found to arise from mostly old (≥10 Gyr) metal-poor ([Fe/H] < -1.0) stars of low-mass progenitor galaxies. New imaging facilities such as LSST, Euclid, and the ``big eyes'' on the sky - ELT and JWST with their advanced instrumentation - promise to greatly increase our knowledge of the progenitors of the IGL and ICL stars, their ages, metal content, masses and evolution, thereby increasing our understanding of this enigmatic component.

## 1. Introduction

The motivation to learn about the physics of extended stellar halos in galaxies and the diffuse light in larger structures is synthesized by the statement of Eggen, Lynden-Bell and Sandage in their 1962 seminal paper (Eggen, Lynden-Bell & Sandage 1962), that is " the time required for stars in the galactic system to exchange their energies and momenta is very long compared with the age of the galaxy. Hence the knowledge of the present energies and momenta of individual objects tells us something of <u>the initial dynamic conditions</u> under which they were formed". This statement from sixty years ago is still very relevant today: we study the dynamics and chemical composition of stars in the diffuse outer halos of galaxies and in the intracluster and intragroup light in clusters and groups, in order to understand the build-up and assembly of these extended structures.

The diffuse stellar component (DSC) in local groups and clusters is an extended, low surface brightness distribution of stars with approximate elliptical symmetry around the bright galaxies at the center of these groups and clusters. The DSC can harbor multiple components from galaxy outer haloes to intracluster light, which can be characterized by their radial profiles, and shapes of the isophotes. Thanks to wide field imaging facilities these DSCs have been measured to extend out to hundreds of kiloparsecs. In addition to the smooth components with elliptical symmetry, maze-like webs of tails, plumes and spurs can be present. Exemplary testimony of such intricate luminous substructures is provided by the extremely deep images, reaching surface brightness levels below 1% of the night sky: see for example those obtained for the Virgo (Mihos 2005, 2013; Spavone et al. 2017), Fornax (Iodice et al. 2016) and the Hydra-I (Arnaboldi et al. 2012; Barbosa et al. 2018; La Marca et al. 2022a, 2022b) clusters.

Attempts to quantify the kinematics of the DSC within the groups and clusters date back to the work of Alan Dressler in 1979, on the dynamics and structure of the cD galaxy IC1101 in the cluster Abell 2029 (Dressler 1979). In his work, Dressler measured absorption lines on the continuum from the stars in spectra obtained for the extended halo around IC1101; the relatively ``bright'' surface brightness of this halo allowed the measurement of the broadening function of the MgII and Fe absorption features around 5200 Å. The width of these absorption lines probes the line-of sight (LOS) velocity dispersion of the stars orbiting in this halo. Dressler found that the measured velocity dispersion increased with central distance from IC1101 implying that the mass-to-light (M/L) ratio of the envelope was increasing rapidly. His assessment of the physical implications of these measurements was that he identified the outermost high M/L component with the stripped stars from satellite galaxies orbiting in the cluster around IC1101. Today we associate such diffuse light around central bright galaxies with the intracluster light (ICL), defined as the stellar component at the centres of groups and clusters which is not bound to individual galaxies.

Kinematic measurements of the stellar motions in extended structures such as the outer halos of galaxies, the ICL and the intragroup light (IGL) are challenging. Absorption line spectroscopy in early-type galaxies (ETGs) is often confined to within two effective radii ($R_e$, defined as the radius which contains half of the total light in a galaxy). Despite the observational challenges, kinematic studies of ETG halos from integrated-light absorption spectra were undertaken by Kelson et al. (2002), Weijmans et al. (2009), Coccato et al. (2010), Murphy et al. (2011), Bender et al. (2015), and Barbosa et al. (2018) using either long slit spectroscopy or integral field spectroscopy (IFS) out to large radii, and integrating for several hours on large telescopes. More recently the SLUGGS survey (Arnold et al. 2014; Foster et al. 2016; Bellstedt et al. 2017), the MASSIVE survey (Raskutti et al. 2014; Veale et al. 2017), Boardman et al. (2017) and Loubser et al. (2022) generated kinematic data from IFSs for larger samples of ETGs. However, the extended kinematics in these studies do not reach distances beyond 3 - 4$R_e$.

The observational limitations hamper the assessment of the complex dynamics of dynamically ``hot'' stellar systems, dominated by velocity dispersion. ``Complex'' here means that the inference of the total enclosed mass from the observed velocities is not straightforward as it is for circular orbits. In hot stellar systems, the total enclosed mass can be constrained using the higher moments of the line of sight velocity distribution (LOSVD). Measuring these higher order moments in galaxy halos may resolve the anisotropy-potential-degeneracy (e.g., Gerhard 1993; Rix et al. 1997; Thomas et al. 2009; Napolitano et al. 2009; Morganti et al. 2013), but beyond 3-4 $R_e$ this is not possible with absorption line spectroscopy.

The way forward to map the kinematics of the DSC to large radii is the spectroscopic follow-up of relatively luminous discrete tracers which overcome the drop in surface brightness. Such discrete tracers are globular clusters (GCs, Schuberth et al. 2010; Strader et al. 2011, Pota et al. 2018 and reference therein) or planetary nebulae. Planetary nebulae (PNe) are established kinematics probes of the stellar populations in extended galaxy halos and ICL (e.g., Arnaboldi et al. 1998, 2005, 2017; Longobardi et al. 2013; Hartke et al. 2017; Pulsoni et al. 2018). Their relatively bright [OIII] 5007 Å line emission stands out against the galaxy faint continuum background, making them relatively easy to detect (Arnaboldi et al. 2002). Since PNe represent the final phase of the Asymptotic Giant Branch (AGB) for low to intermediate mass stars, in old stellar populations such as found in ETGs they trace the bulk of the stars that emit the host-galaxy continuum light. In the overlapping regions, the kinematics of PNe is directly comparable to integrated light measurements (Hui et al. 1995; Arnaboldi et al. 1996, 1998; Méndez et al. 2001; Coccato et al. 2009; Cortesi et al. 2013; Pulsoni et al. 2018).

From the projected two-dimensional kinematics maps derived with discrete tracers like PNe and GCs it becomes possible to map substructures both in space and velocity (Napolitano et al. 2003; Longobardi et al. 2015b; Hartke et al. 2018; Napolitano et al. 2022). In the last few years evidence was collected for debris of disrupted satellites in halos, groups and clusters, providing observational evidence that these components are still building up (Ventimiglia et al. 2011; Longobardi et al. 2015b; Hartke et al. 2018). To understand the growth of these structures, observations must be compared with models and simulations. This is because we deal with a long sequence of events, where accretion and mergers play important roles (see also Arnaboldi & Gerhard 2010; Pulsoni et al. 2021), leading to the formation of the IGL and ICL that we see today in the nearby universe.

Since the first spectroscopic identification of intracluster planetary nebulae (ICPNe) in the Virgo cluster (Arnaboldi et al. 1996), imaging surveys and extended spectroscopic follow-up campaigns have identified these kinematic probes and used them to trace the radial extent and the kinematics of the DSC in groups and clusters in the nearby universe. PNe as discrete tracers also play an important role for galaxy halos, in order to constrain the integrated mass as function of radius, the intrinsic three-dimensional shape of the mass distribution, luminous and dark, and the orbital distribution of halo stars in the nearby universe. See for example the recent comparisons between the observed physical properties of ETGs outer halos (Pulsoni et al. 2018) with their analogs selected from the Illustris TNG simulations (e.g. Pulsoni et al. 2020, 2021).

In this article we describe the IGL and ICL properties in the local groups[1] and clusters, as mapped by the kinematics and spatial distribution of their stars[2]. We begin with a concise overview of the general properties of the IGL and ICL from deep imaging and quantitative photometry, to then move on to utilize them together with the kinematics measured via absorption lines on the stellar continuum and/or PN spectroscopy. Important guidance comes from numerical simulations and semi-analytic models, which are used as abenchmark to identify the most robust observables in the group and cluster, but also as a tool to interpret the results of the observations. We then explore ways to investigate observationally the stellar populations of the IGL and ICL, to constrain their progenitors, and compare the measured kinematics of distinct discrete tracers, such as PNe and GCs. We then draw our conclusions and illustrate avenues for future investigations.

---

[1] The closest compact group is at D> 100 Mpc; for a summary of the properties of IGL in compact groups see Poliakov et al. (2021)

[2] We do not discuss the Intra Cluster Medium properties, either hot gas or dust

## 2. IGL and ICL properties in local groups and clusters: their light distribution

The properties of IGL and ICL in groups and clusters measured from deep imaging and surface brightness photometry are the main topics of several other contributions in this journal. In this review, we attempt to collect the basic facts which set the stage for an overview of the IGL and ICL kinematics in local groups and clusters. The presence of ICL is historically assessed as an excess of light with respect to the de Vaucouleurs $R^{1/4}$ profile (de Vaucouleurs 1948), extrapolated from the galaxy at the cluster center to large cluster radii (Thuan & Kormendy 1977; Melnick et al. 1977; Schombert 1986; Bernstein et al. 1995). The central group or cluster galaxy (CCG) is often but not always the brightest cluster galaxy (BCG) and often but not always a cD galaxy, i.e. a giant elliptical with cluster-size extra light at large radii with respect to the outward extrapolation of a Sersic (1963) function fitted to the inner profile (Kormendy et al. 2009).

Because of this light excess, either one or more components modeled as Sersic or exponential profiles are commonly used to reproduce the surface brightness profile over the entire range of radii, from a few to hundreds of kiloparsecs (Gonzalez et al. 2005; Spavone et al. 2017). These surface brightness profiles have a range of radial gradients, from steeper to flatter slopes across different clusters (Krick & Berstein 2007; Kluge et al. 2020). The use of several components with different analytical formulae to reproduce the surface brightness profile from the inner to the outer regions is further motivated by a frequent change of the isophote shapes with radius. The ellipticity generally increases with radius, i.e. isophotes become flatter, and the position angle profile of the major axis of the outer isophotes has sharp variations (Gonzalez et al. 2005; Kluge et al. 2021) as function of radius outwards from the CCG to the group and cluster regions.

This behavior signals the transition from the CCG halo to the IGL and ICL. In general, the ICL is more aligned with the host cluster than the CCG in terms of position angle, ellipticity, and centering (Zibetti et al. 2005; Kluge et al 2021). The quantitative 2D morphological variations (flattening of the isophotes, twist of the positional angle of the major axis with radius) and photometric properties (different radial gradients in different clusters) suggest that the IGL and ICL are dynamically young and separate from the CCG halo.

When mapping the DSC in groups and clusters on the basis of photometry, the question arises whether the separation of the IGL and ICL from the outer halo of the CCG has a physical motivation or whether it is only a taxonomy exercise, focused primarily on assessing changes of the radial gradients or isophotal twists of the surface brightness distribution. The latter approach which unifies both ICL and the halo of the CCG as DSC can be summarized by the statement of Uson et al. (1991): "*Whether this diffuse light is called the cD envelope or diffuse intergalactic light is a matter of semantics; it is a diffuse component which is distributed with elliptical symmetry about the center of the cluster potential*". This sets the stage for a statistical study of the CCG and ICL by stacking deep images of clusters after suitable rescaling, such as utilized by Zibetti et al. (2005), see also D'Souza et al. (2014). Zibetti et al. (2005) studied the surface brightness profiles of the DSC components obtained after stacking images of 683 clusters in the redshift range 0.2 - 0.3, from the first data release of the Sloan Digital Sky Survey. They were able to detect ICL out to 700 kpc from the CCGs as excess with respect to an inner $R^{1/4}$ profile; they determined an ICL faction of 10.9%±5%, which seems to be independent of cluster richness and CCG luminosity. While this ``global" approach provides the average distribution of the DSC, it erases any differences of shapes and position angles as function of radius.

Investigations of the properties of the DSC in clusters from cosmological simulations can be carried out with a very similar methodology, see discussions in Cooper et al. (2013, 2015) for particle tagging N-body simulations and in Pillepich et al. (2018), Cañas et al. (2020) for fully hydrodynamical cosmological simulations. In simulations, the stellar particles which remain in a group or cluster once all the sub-halos or satellite galaxies have been removed using binding energy constraints, can be readily interpreted as the close analog of the observed DSC in these structures, including the CCG. From the studies based on the Illustris TNG simulations, Pillepich et al. (2018) obtained that the radial profiles of the DSC at large radii become shallower in the more massive clusters.

A complementary approach to that described by Uson et al. (1991) and in the previous paragraphs seeks to associate major changes in the spatial distribution and surface brightness (radial gradient, ellipticity or position angle profiles) of the DSC in groups and clusters with the transition to a different physical component, e.g. the IGL or ICL. This strategy is similar to that carried out for the structural components in galaxies (bulge versus disk decomposition, see Kormendy & Bender 2012).

The fraction of the IGL and ICL in individual groups and clusters is then derived from quantitative photometry. The estimated values depend also on the methodology used to quantify them, i.e. whether the ICL is measured as the light above a fixed surface brightness threshold, or computed as the light excess with respect to the extrapolation of the $R^{1/4}$ law, or as fraction of the light associated with a second, outer Sersic component (see extensive comparisons and discussions in Kluge et al. 2021 and Montes 2022). It also depends on the radial range over which this component is measured. Typically measuring light above a certain threshold provides the largest values of the IGL and ICL fractions with respect to the total galaxy light, which is determined by summing up all the light in the CCG and satellite galaxies. Fractions obtained from the different methodologies vary from few percent in the low mass groups (4% in the Leo I group, Watkins et al. 2014; Hartke et al. 2020; 10% in the Virgo subcluster B, see Hartke et al. 2018 ) and unrelaxed clusters ( e.g. 5% in Virgo, see Castro-Rodriguez et al. 2009),  up to an average 48-52% in the more relaxed, dynamically evolved clusters (computed as excess to the extrapolated $R^{1/4}$  profile or associated with the second Sersic component, Kluge et al. 2021).  See also Cañas et al. (2020) and Contini & Gu (2020) for predictions on the IGL and ICL fractions in groups and clusters from cosmological simulations. These recent simulations reproduce the increase of the amount of light at 100 kpc with respect to the inner 10 kpc in groups and clusters as function of mass, measured by DeMaio et al. (2020).

An important comparison is that between the surface brightness profile measured for the ICL, and the surface brightness profile obtained from the averaged light of all group and cluster satellite galaxies. Where it has been measured, the ICL surface brightness is steeper than that of the averaged-in-annuli (group or cluster) satellite galaxies surface brightness profile, i.e. the ICL more spatially concentrated to the high density regions (Zibetti et al. 2005; Castro-Rodriguez et al. 2009; Spiniello et al. 2018). For more exhaustive reviews of these results see the contributions in this volume focused on the quantitative photometric properties of the IGL and ICL.

### 3. DSC in groups and clusters: how to kinematically disentangle IGL and ICL from the outskirts of the CCG

Another approach to identify IGL and ICL is via measurements of the kinematic of stars. This is also motivated by the observed rise of the velocity dispersion in CCGs, which started with the work of Dressler (1979) on IC 1101. This early work signaled a transition from stellar motions bound to the CCG to motions gravitationally bound to the group or cluster structure as a whole. It was then the first kinematic detections of IC PNe in the core of the Virgo cluster (Arnaboldi et al. 1996) and in Coma (Gerhard et al. 2005, 2007), which displayed large residual velocities with respect to the closest galaxies (>1600 kms$^{-1}$ relative to M86 in Virgo and ~700 kms$^{-1}$ relative to NGC 4874 in Coma) that provided the opportunity to distinguish CCG halo and IGL or ICL at the same location. This early work was subsequently expanded to provide LOSVDs and velocity histograms of stellar tracers on larger areas: see Arnaboldi et al. (2004), Doherty et al. (2009) for the Virgo cluster and Ventimiglia et al. (2011) for the Hydra I cluster. When LOS velocities are measured to be significantly in excess of the velocity dispersion measured in the CCG, then the gravitational potential of the entire group or cluster must provide the pull to bind these high velocity tracers, at the location where they are detected.

The kinematically motivated approach to separate the CCG halo from ICL was adopted by Dolag et al. (2010) and Cui et al. (2014) for simulated clusters, using the distribution[3] of absolute velocities of stellar particles. In their cosmological cluster simulations, these authors found that the velocity distribution of the stellar particles associated with the CCG and DSC (after all the stellar particles in the satellite galaxies are removed) is bimodal in the volume occupied by a given cluster and can be fitted by two Maxwell-Boltzmann distributions, see Figure 1. The ``narrower'', i.e. colder, particle distribution occupies a more limited region of the phase space; these particles are associated with the central galaxy and the mass profile is steeper with radius, see Figure 1. The second ``broader'', i.e. dynamically ``hotter'', particle distribution is more extended in the cluster volume, it has a shallower mass density profile and it is thus responsible for the ``light excess'' at large cluster-centric radii. The physical properties of these two components of the DSC, correspond to the CCG and ICL stars respectively. These results from simulations provide a consistent framework for the interpretation of the observed increase in the velocity dispersion profiles around IC 1101 (Dressler 1979), M87 (Murphy et al. 2011, Longobardi et al. 2015a), NGC 6166 (Bender et al. 2015), and NGC 3311 (Arnaboldi et al. 2012, Barbosa et al. 2018), whereby the LOSVD for the DSC in these clusters can be considered as the sum of the ``colder'' CCG halo and the ``hotter'' ICL. We shall discuss these observational results more in detail in the next sections.

The physical criterium which is applied to kinematically separate the ICL from the CCG in these simulated clusters is that of the binding energy, i.e. whether the observed velocity component would satisfy the Virial theorem for the mass associated with the closest galaxy. By adopting the physical criteria based on binding energy, we can then identify the ICL stellar particles; whether this may always be possible as function of decreasing group and cluster masses, and dynamical status of the group or cluster, would require further analysis of cosmological simulations.

---

[3] Histogram of the modulus of velocities

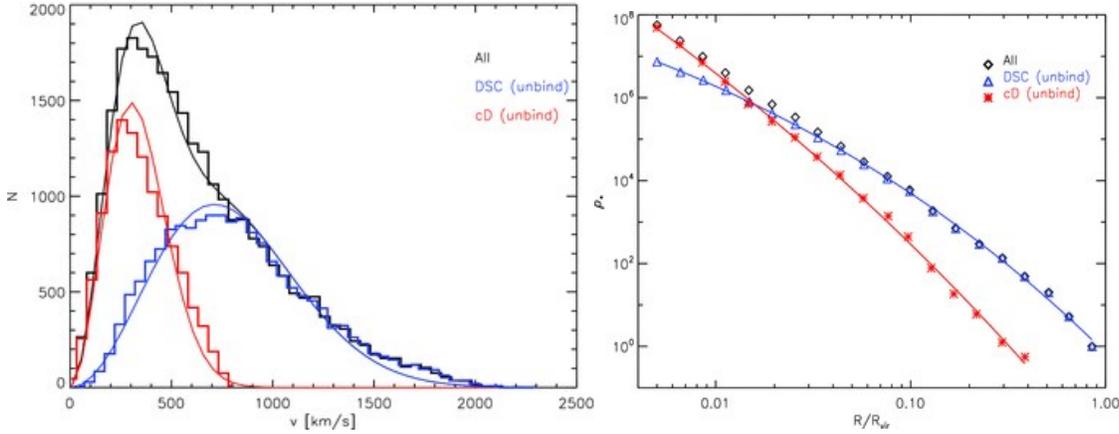

**Figure 1:** Maxwell-Boltzmann distributions of the absolute velocities of stellar particles in simulated clusters and separation of central galaxies and ICL. Left panel - velocity histogram (black line) of the DSC in a simulated cluster and a double Maxwellian fit to it (grey line) are shown. Red and blue histograms show the velocity distribution of the CCG and ICL stars after the unbinding procedure, together with the two individual Maxwellian distributions from the global fit, respectively. Right panel – All stars centered on CCG (black symbols), CCG (red symbols) and ICL (blue symbols) stellar density profiles. The red and blue line are Sersic fits to the two distributions, respectively. Image reproduced from Dolag et al. (2010) with permission.

One can further explore the properties of the kinematically selected ICL particles in simulations. It turns out that the ICL stellar particles in hydro-dynamical simulations selected in the aforementioned way have systematically different star formation histories (Kapferer et al. 2010) and are on average older that the stars in cluster galaxies (Murante et al. 2004) in addition to their distinct phase space distribution. This is a very interesting perspective for the investigation of ICL in clusters. It provides a complementary, potentially more powerful approach to the statistical one described by Uson et al. (1991), which seeks to unify CCG halo and ICL in a unique DSC. The results from simulations stimulate the need for observational strategies to measure these differences and investigate how they were established during the galaxy assembly and cluster formation. *Mapping differences in kinematics and stellar population properties thus provides the opportunity to tag CCG halo and ICL stars at the same spatial location.* In sections 4, 5 and 6 we review highlights and examples where it has been possible to observationally distinguish CCG halo, IGL and ICL via kinematic measurements and to link the distinct kinematics to differences in surface brightness distributions and stellar populations, which in turn leads to constraining their origins in a quantitative way, in terms of progenitor masses and timing of the disruption events.

**4. CCG and IGL or ICL as distinct kinematics components**

The measurement of the surface brightness profiles and of detailed kinematics at large radii from the CCG center would reveal the presence of the accretion substructures that are not yet phase mixed, as well as diffuse ICL components. These measurements would allow one to isolate the smooth CCG halo, constrain its orbit distribution and obtain unbiased estimates of its stellar and enclosed dynamical mass. By smooth CCG halo here we refer to the part of the halo that is approximately phase-mixed, i.e. has approximately Gaussian LOSVD centered about the systemic velocity of the galaxy. This is clearly different from a multi-peaked LOSVD. The working concept of a smooth halo is useful because the approximately Gaussian LOSVD enables one to determine a well-defined halo velocity dispersion profile and combine it with the tracer density to carry out a Jeans analysis of the mass distribution at large radii. The use of these concepts, i.e. smooth CCG halo and less well mixed dynamically hotter IGL and ICL allows one to investigate the transition between the CCG halo and the IGL or ICL.

Since the serendipitous discovery of three ICPNe at $V_{LOS,mean} = 1400$ kms$^{-1}$ along the LOS of the galaxy M86 (NGC 4406; $v_{sys} = -240$ kms$^{-1}$) in the Virgo cluster core (Arnaboldi et al. 1996), ICPNe have been used to map both spatial distributions and kinematics of the ICL and IGL component in nearby clusters. In what follows, the kinematic properties of the ICPN samples in the clusters within 100 Mpc distance, i.e. Virgo, Fornax, Hydra I and Coma, are outlined and compared, including efforts to physically separate the CCG halo from the cluster ICL or group IGL component.

## 4.1 Clusters in the nearby universe

**Kinematics of ICL around M87 in the Virgo cluster (D=15 Mpc)** - Narrow band imaging surveys carried out to identify ICPNe were performed by Feldmeier et al. (1998, 2004), Arnaboldi et al. (2002), Okamura et al. (2002), Aguerri et al. (2005), Castro-Rodriguez et al. (2009), Longobardi et al. (2013). These were followed up by efforts to obtain spectra for ICPNe in the Virgo cluster (Arnaboldi et al. 2003, 2004; Doherty et al. 2009), which provided reliable measurements for 52 single PNe within one degree circular radius around M87 and in the Virgo core (as defined by Binggelli et al. 1987). The spectra for these single stars showed the [OIII] λ4959/5007 Å double emissions, confirming the identification of these emission line sources as PNe and their physical association with the Virgo cluster. The sparsely sampled projected phase-space, built from the ICPN $V_{LOS}$ versus radial distance from the centre of M87, displayed regions with different kinematics. The halo of M87 was clearly identified as a single, relatively narrow ridge centred around the galaxy systemic velocity ($V_{sys,M87} \approx 1300$ kms$^{-1}$). However, the data showed an additional component to the M87 halo, which was characterised by a broad and extended LOSVD out to distances of 2000 arcsec, ≈150 kpc, and larger. Such a broad LOSVD is reminiscent of the velocity distribution of the satellite galaxies in the Virgo cluster (Binggeli et al. 1987).

An important step in mapping the IC component around M87 came from the more recent, extended, ICPN imaging and spectroscopic survey carried out by Longobardi et al. (2013, 2015a). This work led to a major increase of a factor 15 in the number of measured $v_{LOS}$ from independent spectra, with respect with previous ICPNe samples in the centre of the Virgo subcluster A (Binggeli et al. 1987). The ICPN survey carried out with Subaru SuprimeCam (imaging) and Flames at UT2 on the VLT (spectroscopy) delivered 287 spectra for individual PNe (Longobardi et al. 2015a). This sample was analysed including additional 11 $v_{LOS}$ from independent PN spectra measured by Doherty et al. (2009) in the same area, giving a combined total sample of 298 PN $v_{LOS}$. Interestingly, the entire PN LOSVD displays broad asymmetric wings that cannot be matched by a single Gaussian, see Figure 2. Instead it can be approximated by two Gaussians, a narrow component centred at the systemic velocity of M87, with $\sigma_{narrow} \approx 300$ kms$^{-1}$, and an additional, much broader, Gaussian with an average velocity at $v_{broad} \approx 1000$ kms$^{-1}$ and $\sigma_{broad} \approx 900$ kms$^{-1}$. Each PN in the projected phase-space, $V_{LOS}$ vs. $R_{major}$, around M87 can be assigned a probability to belong either to the narrow Gaussian, i.e. the M87 halo component, or to the broad component, i.e. the ICL. The ICPNe in the Virgo cluster core were thus identified by Longobardi et al. (2015a) via a multi-Gaussian decomposition of their LOSVD. The ICPN sample was later expanded by Longobardi et al. (2018a), who added the unrelaxed component that overlaps with the M87 smooth halo in the 40-150 kpc radial range. Using the combined ICPN sample, i.e. ICL from the broad Gaussian ($\sigma_{broad} \approx 900$ kms$^{-1}$) in the Virgo core plus the unrelaxed components at the location of the M87 halo (including the crown, Longobardi et al. 2015b), the resulting ICPN LOSVD is statistically undistinguishable from the LOSVD for satellite galaxies (Longobardi et al. 2018a) in the Virgo core region (Binggeli et al. 1993). This approach is the implementation in the observable parameter space of the CCG halo/ICL decomposition of the DSC in simulated clusters, as illustrated by Dolag et al. (2010). For the GCs populations in the Virgo core, the separation between a GCs population bound to M87 and a GCs population orbiting in the IC space was determined independently by Longobardi et al. (2018b).

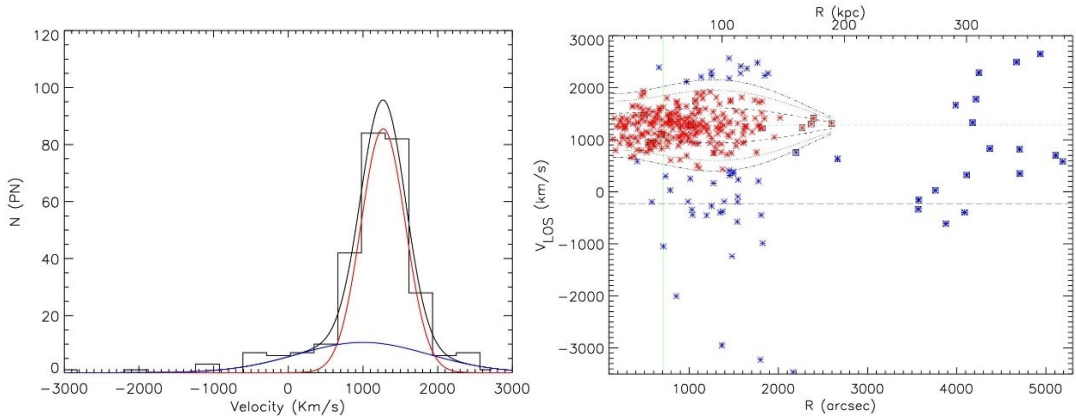

**Figure 2:** PN velocities tracing the diffuse light around M87 in the Virgo cluster. Left panel - Histogram of the $V_{LOS}$ of the spectroscopically confirmed PNe (black histogram) fitted with a double Gaussian (black curve). Red and blue lines represent the two Gaussians associated with the M 87 halo and the IC components. Right panel - Projected phase-space diagram, $V_{LOS}$ vs. major axis distance from the centre of M 87, for all spectroscopically confirmed PNe in Longobardi et al. (2015a). The major axis distance is given both in arcsec (bottom axis), and in kpc (top axis), where 73 pc = 1″. The PNe are classified as M 87 halo PNe (red asterisks) and ICPNe (blue asterisks), respectively; see text. Black squares identify spectroscopically confirmed PNe from Doherty et al. (2009). The smoothed 1, 2, and 2.5σ thresholds are represented by the dashed, dotted, and dot-dashed lines, respectively. The dashed horizontal line shows the M 87 systemic velocity $V_{sys}$ = 1275 kms$^{-1}$, while the continuous green vertical line shows the effective radius $R_e$ = 704″ determined by Kormendy et al. (2009). At $V_{LOS}$ = −220 kms$^{-1}$, we plot the M 86 systemic velocity (long dashed line). Image reproduced from Longobardi et al. (2015a) with permission.

**Kinematics of the ICL around NGC 1399 in the Fornax cluster (D=18 Mpc) –** The extended halos of NGC 1399, NGC 1316 and the ICL in the Fornax cluster have been targets of several extended imaging and spectroscopic surveys to measures the $V_{LOS}$ of discrete tracers including PNe (Arnaboldi et al. 1994; Mc Neil et al. 2010, 2012; Spiniello et al. 2018) and GCs (Schuberth et al. 2010; Pota et al. 2018; Chaturvedi et al. 2022). For the earlier imaging survey to identify ICPNe in Fornax see also Theuns & Warren (1997). The outer envelope of NGC 1399 is characterized photometrically by an exponential profile (Iodice et al. 2016 and reference therein) which dominates the DSC in Fornax out to 200 kpc. Once the different tracers are merged, one wishes to probe the kinematics from the center, using absorption line spectra, then out to 200 kpc using PNe, red and blue GCs. The combined velocity dispersion profile shows a central velocity dispersion value of ∼ 380 kms$^{−1}$, which then decreases rapidly to 250 – 200 kms$^{−1}$ between 10 – 20 kpc radius. Beyond those radii, $\sigma_{LOS}$ increases again reaching 350 kms$^{−1}$ at 50 kpc and then decreases to 300 - 250 kms$^{−1}$ at larger radii. The $\sigma_{LOS}$ values of the DSC in the Fornax cluster at ∼ 200 kpc from NGC 1399 are lower than the velocity dispersion measured from the $V_{LOS}$ of the satellite galaxies at those radii. Figure 3 shows a complete mapping of the velocity dispersion profile in the Fornax cluster from the different tracers and satellite galaxies.

The sample of PNe (Spiniello et al. 2018) and GCs (Pota et al. 2018) was used by Napolitano et al. (2022) to identify substructures in the Fornax cluster projected phase space, for which they provide a catalogue of 13 stellar stream candidates, each with their mean centroid position, associated number of PN and GC tracers, radial velocity, size, total luminosity and surface brightness. Most of these streams are kinematically connected to ultra-compact dwarfs (UCDs), which supports a scenario of their morphological transformation from the dwarf progenitors, after being disrupted in the cluster gravitational field. Some of the additional long coherent substructures connecting cluster members and some isolated clumps of tracers not associated with UCDs may represent fossil records of satellites that have since merged with Fornax cluster member galaxies.

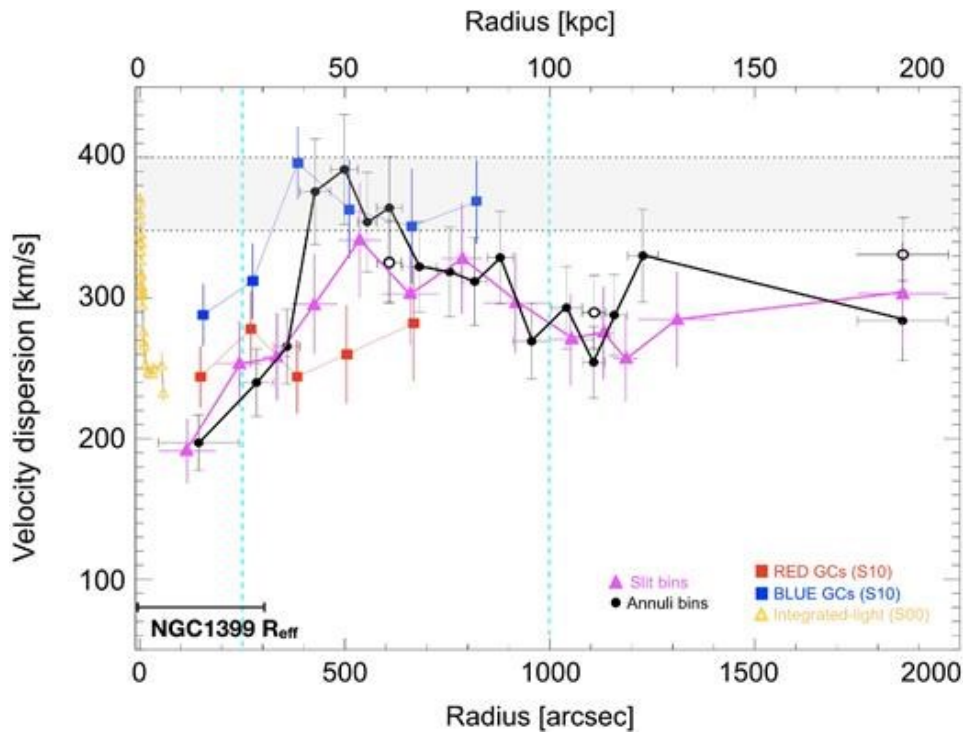

**Figure 3:** Velocity dispersion profile of diffuse light in the Fornax cluster, from the center of NGC 1399 out to 200 kpc. The $\sigma_{LOS}(R)$ from PNe is plotted as filled black circles. The three empty symbols show values at radii where PNe most likely associated with satellite galaxies were removed. $\sigma_{LOS}(R)$ from red and blue GCs (see Schuberth et al. 2010; S10 in the figure) are shown as red and blue squares, respectively. The inner velocity dispersion profile from absorption line spectra (Saglia et al. 2000; S00 in the figure) is shown as yellow triangles. The grey shaded horizontal region represents the $\sigma_{LOS, gal}$ of the Fornax cluster with its 1σ error. Image reproduced from Spiniello et al. (2018) with permission.

**Kinematics of ICL around NGC 3311 in the Hydra I cluster (D=50 Mpc) -** With an optimized observational technique named Multi Slit Imaging Spectroscopy (MSIS: Gerhard et al. 2005; Arnaboldi et al. 2007) with FORS2 at the ESO VLT, Ventimiglia et al. (2011) detected a total of 56 PNe in a single field of 100×100 kpc$^2$, centered on the galaxy NGC 3311, in the core of the Hydra I cluster. The PN LOSVD in this region has several peaks. In addition to a broad symmetric component centered at the systemic velocity of the cluster ($V_{Hydra}$ = 3683 kms$^{-1}$; $\sigma_{Hydra}$ = 784 kms$^{-1}$), two narrow peaks are detected at 1800 kms$^{-1}$ and 5000 kms$^{-1}$. We address the origin of the broad component and the secondary peaks in turn. In a series of papers, Barbosa et al. (2016, 2018) investigated the velocity dispersion profile of the galaxy NGC 3311 along its photometric major axis, and carried out a physical modelling of both the photometry and projected kinematics on the sky from deep IFS observations with MUSE at the ESO VLT. They found that the light in this galaxy can be modelled as a superposition of several non-concentric spheroidal components with increasing scale, slightly different systemic velocities and velocity dispersions. Interestingly, the outermost envelope is offset with respect to the central region by up to 50" to the North. They found that the central high surface brightness components have a relative motion and displacement from the outer envelope, whose velocity dispersion rises to cluster-core values of ~ 400-500 kms$^{-1}$ already at R> 20". This outer envelope has an exponential surface brightness profile and is dynamically associated with the cluster (Arnaboldi et al. 2012). The transition to cluster-like velocity dispersion around NGC 3311 is confirmed independently by the LOS velocities of GCs (Hilker et al. 2018). The entire inner region is further displaced from the Hydra I cluster's center, as defined by the spatial distribution of the outer satellite galaxies, presumably due to a recent sub-cluster merger (Barbosa et al. 2018).

The presence of secondary peaks in the LOSVD of the PNe indicates the existence of un-mixed populations in the Hydra I cluster core. The spatial distribution of the PNe associated with the narrow velocity sub-component at 5000 kms$^{-1}$ is spatially located on an excess of light in the North-East quadrant of NGC 3311, in the B band surface photometry (Arnaboldi et al. 2012). At the same sky area, a group of dwarf galaxies with $V_{LOS} \approx 5000$ kms$^{-1}$ is detected. Deep long-slit spectra at the position of the dwarf galaxy HCC 026 at the center of this light excess show absorption features whose observed wavelengths are consistent with a $V_{LOS} \sim 5000$ kms$^{-1}$. Arnaboldi et al. (2012) concluded that the PNe associated with the LOSVD peak at $\approx$5000 kms$^{-1}$, spatially overlapping with the dwarfs and the light excess in the North-East quadrant of NGC 3311 are physically associated. Given the large residual velocity ($\sim 1400$ kms$^{-1}$) from the systemic velocity of NGC 3311, they are part of the ICL. About the secondary velocity peak at ~1800 kms$^{-1}$: within a 20 arcmin distance from NGC 3311 there are 8 galaxies with systemic velocities lower than 2800 kms$^{-1}$ and R band total magnitude < 15.4. Two of these galaxies are spirals, with NGC 3312 being at the boundary of the MSIS regions. These ICPNe may therefore be associated with the infalling group associated with NGC 3312. Hence also in the case of the relaxed Hydra-I cluster, PN kinematics, photometry and deep absorption spectra support the evidence for accretion or infalling of groups, including a sub-cluster merger, with a significant fraction of stars being added to the ICL around NGC 3311.

**Kinematics of ICL in the core of the Coma cluster (D=100 Mpc) -** The Coma cluster is the richest and densest of the nearby clusters, yet there is evidence that sub-cluster mergers are taking place. With the MSIS technique using the FOCAS spectrograph at the Subaru telescope, Gerhard et al. (2005) detected the [OIII] 5007 Å emission for ICPNe and the $V_{LOS}$ for 37 of them were measured (Arnaboldi et al. 2007). These ICPNe are associated with the diffuse light previously detected by Bernstein et al. (1995) near the sky position α(J2000) 12:59:41.784; δ(J2000) 27:53:25.388. These ICPNe in Coma are the most distant single stars whose spectra can be measured with 8-meter class telescopes, apart from cosmological supernovae. Gerhard et al. (2007) detected several peaks in the LOSVD of the ICPNe within a 6' diameter field, ~5' away from the cD galaxy NGC 4874. A velocity sub-structure is present at $\sim 5000$ kms$^{-1}$, probably from in-fall of a galaxy group (Adami et al. 2005), while the main IC stellar component has a $V_{mean}$ centered around ~6500 kms$^{-1}$, about $\Delta v \approx -700$ kms$^{-1}$ offset from the nearest cD galaxy, NGC4874 ($v_{sys} = 7224$ kms$^{-1}$). The ICPN $V_{mean}$ is consistent with the systemic velocity of the other D galaxy in the Coma core, NGC 4889. The result of these observations connects the stars and their motions at the location of the MSIS field with a thick plume or tail stripped off NGC 4889 during an on-going binary sub-cluster merger in the Coma core (Fitchett & Webster 1987; Colless & Dunn 1996). Both the kinematics and elongated morphology of the ICL in Coma (see Thuan & Kormendy 1977) show that the cluster core is in a highly dynamically evolving state; this is independently supported by galaxy redshift measurements (Adami et al. 2005) and X-ray data (Neumann et al. 2003). The sub-cluster associated with NGC 4889 is likely to have fallen into Coma from the eastern A2199 filament, in a direction nearly in the plane of the sky, meeting the NGC 4874 sub-cluster arriving from the west. The two inner sub-cluster cores are presently beyond their first and second close passage, during which the elongated distribution of diffuse light was created, see Figure 4. The absence of a cooling core in the Coma cluster is consistent with near head-on collision of the two sub-clusters. Gerhard et al. (2007) also argued that the extended western X-ray arc (see Neumann et al. 2003) traces the arc shock generated by the collision between the two sub-cluster gaseous halos. The high-density mass concentrations associated with the sub-clusters around NGC 4874 and NGC 4889 in the Coma cluster's central region were confirmed independently by weak lensing measurements (Okabe et al. 2010).

**Summary kinematics of ICL in local clusters –** Kinematics measurements from discrete tracers clearly support the ICL to be a distinct physical component from the CCG halo, which can be offset spatially from the central CCG. The ICL LOSVD has i) width approaching values expected from the cluster velocity dispersion and ii) may have additional multiple peaks. Those multiple peaks, when present, are related with sub-clusters and their on-going mergers, see for example the binary mergers in the central region of the massive Coma cluster, or the displaced/offset envelope surrounding NGC 3311. Kinematics substructures signaled by independent peaks in the ICL LOSVD correlate with the presence of photometric substructures and satellites at similar LOS velocities.

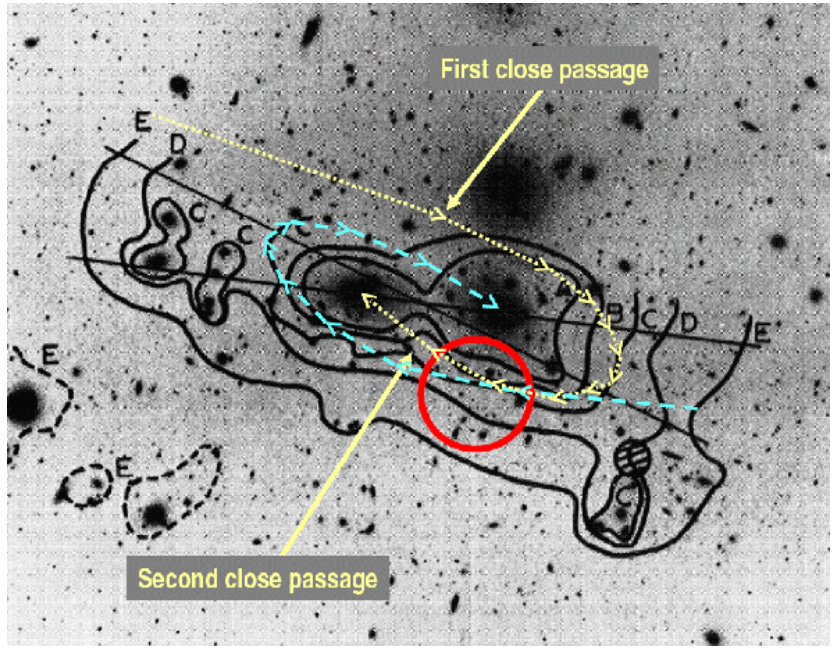

**Figure 4:** Diffuse ICL in the Coma cluster core. The position of the MSIS field studied in Gerhard et al. (2005, 2007) is indicated by the red circle. It is superposed on the diffuse light isodensity contours drawn by Thuan & Kormendy (1977). The MSIS field is about 5′ south of NGC 4874. The second Coma cD galaxy NGC 4889 is 7′ east (to the left) of NGC 4874. The bright object north of NGC 4874 is a star. Note the strong elongation of the ICL distribution in the Coma cluster core. The likely orbits of NGC 4889 and NGC 4874 up to their present positions are sketched by the yellow dotted and magenta dashed lines, respectively; image reproduced from Gerhard et al. (2007) with permission.

### 4.2 Properties of the IGL in local galaxy groups

In this section we present a concise overview of the kinematics of the IGL in nearby local groups. We shall look at Virgo sub-cluster B with its CCG, M49, and the Leo I group with M105.

**Kinematics of IGL around M49 in the Virgo sub-cluster B (D=15 Mpc)** - M49 (NGC 4472) is the BCG in the Virgo sub-cluster B (Bingelli et al. 1987). Because of its smooth X-ray halo (Böhringer et al. 1994), the temperature map around M49 can be used to trace the gravitational potential within 30 kpc (Das et al. 2010) and thus alleviate the anisotropy-potential degeneracy. From the extended Planetary Nebula Spectrograph (PN.S) ETG survey (Douglas et al. 2007; Arnaboldi et al. 2017; Pulsoni et al. 2018) and the deep photometric sample obtained with the Subaru SuprimeCam by Hartke et al. (2017), a cumulative PN catalogue was assembled to map the kinematic properties out to 95 kpc radius. Hartke et al. (2017, 2018) implemented a multi-Gaussian analysis for the LOSVD in order to identify sub-populations with distinct kinematics, which showed the presence of three kinematically distinct components around M49 (Hartke et al. 2018). These authors found i) a kinematic substructure associated with the on-going accretion of the dwarf irregular galaxy VCC 1249 onto the M49 halo. The two additional kinematically distinct PN populations found around M49 are ii) the main M49 halo and ii) the IGL, respectively. These two components have velocity dispersions $\sigma_{halo} \simeq 170$ kms$^{-1}$ and $\sigma_{IGL} \simeq 400$ kms$^{-1}$ in the 10-80 kpc radial range, respectively. The kinematically selected PNe which are part of the M49 halo have different stellar population properties from the intra group (IG) PNe which we shall present in the following sections. The velocity dispersion $\sigma_{IGL}$ of the IGPN profile at 80 kpc radial distance from the M49 center is consistent with a flat circular velocity curve extended outwards from the region with the X-ray observations. The $\sigma_{IGL}$ measured profile joins onto that of the satellite galaxies in the Virgo sub-cluster B at 100 kpc radius. This was the first time that the transition from a CCG halo to the surrounding IGL is observed based on the velocities of individual stars. For M49 this transition was later confirmed independently from the LOSVD of blue GCs (Taylor et al. 2021).

**Kinematics of IGL around M105 in the Leo I group (D=10 Mpc)** - M105 (NGC 3379) is an ETG in the nearby Leo I group. While classified as a poor group in a low mass environment, the Leo I group is the closest which contains all galaxy types. It thus allows one to explore the physical properties of IGL at the low-mass end of the group mass function. Hartke et al. (2020, 2022) present an extended, imaging and kinematic survey of PNe in M105 and its surrounding, covering 0.25 deg$^2$ area in the Leo I group with SuprimeCam Narrow band imaging and the PN.S. For previous results from a narrow band survey in [OIII] centered on the HI gas cloud in Leo I group see Castro et al. (2003). Hartke et al. (2022) employ Gaussian mixture models to first separate those PNe that are gravitationally bound to the companion galaxy NGC 3384. Then the M105 PNe (169 in total) are further classified as belonging either to the galaxy's inner Sersic halo or to the surrounding outer exponential envelope, which is traced by an excess of metal poor stars (see Hartke et al. 2020; for reference to the HST observations and analysis see Harris et al. 2007; Lee & Jang 2016). Hartke et al. (2022) constructed the smooth velocity and velocity dispersion fields, and determined projected rotation, velocity dispersion, and $\lambda_R$ profile for each component. The M105 halo PNe exhibit a rapidly declining velocity dispersion from $\sigma_0 \sim 220$ kms$^{-1}$ at the center to as low as 50 kms$^{-1}$ at about 6R$_e$, with R$_e$ = 57" (Hartke et al. 2022). Such a negative gradient of the velocity dispersion profile is steeper than that predicted by the R$^{-1/2}$ Keplerian decrease (Romanowsky et al. 2003; De Lorenzi et al. 2009). Differently from the PN subsample linked to the inner Sersic component, the velocity dispersion and rotation traced by the PNe associated with the outer exponential envelope around M105 remain approximately constant at $\sigma_{exp} \sim 160$ kms$^{-1}$ and V$_{rot}$=75 kms$^{-1}$ at 12 R$_e$. These velocity dispersion and rotation values traced by the PNe associated with the exponential envelope are consistent with those measured for the dwarf satellite galaxies in the Leo I group. This indicates that the exponential envelope traces the IGL in the Leo I group. The consistency between the kinematic transition from M105 halo and IGL in the Leo I group and photometric components has been recently confirmed independently by the deep photometry carried out with the VST by Ragusa et al. (2022).

**Summary kinematics of IGL in local groups** – The kinematics observations of the IGL support that the stars in these components originate from the group satellite galaxies. IGL stars are accumulated around the central group galaxies. The IGL in local groups is more relaxed and phase mixed than ICL in clusters. The kinematically selected IGL stars follow exponential profiles away from the central galaxies of these groups.

## 5 Constraints on the orbits of outer halo, IGL, and ICL stars, and their dynamical origin

From their kinematic properties discussed in the previous sections, several components such as the ICPNe in Virgo or Coma appear far from dynamical equilibrium. Others such as the IGL stars around M49 and M105 or the ICPNe in Fornax have smoother photometric and kinematic properties, suggesting that they are more well-mixed in the gravitational potential. In this section we discuss for several of these populations the constraints that can be obtained on their orbits, their accretion history, or the mass distributions in their host systems. We begin by reviewing some general results that we draw on.

**5.1 Some results from dynamical modelling and cosmological simulations**

From dynamical modelling (e.g., Gerhard et al. 2001; Poci et al. 2017) and strong lensing analysis (e.g., Koopmans et al. 2009; Li et al. 2019) the inner high surface brightness regions of massive ETGs have near-isothermal total mass density profiles, i.e., power laws $\rho \propto R^{-\gamma}$ with $\gamma \approx 2.0$-$2.2$. To measure the total mass distribution from stellar kinematics at radii where dark matter contributes substantially, requires information on the shape of the stellar LOSVD, often parametrized through its third (h3) and fourth (h4) and possibly higher-order Gauss-Hermite moments (Gerhard 1993; van der Marel & Franx 1993). The parameter h3 is proportional to the skewness and measures the asymmetric deviations of the LOSVD from a Gaussian, often related to rotation when the low-velocity tail of distribution dominates. The parameter h4 is proportional to the kurtosis of the LOSVD; it probes symmetric deviations of the LOSVD, indicating either a more peaked (h4 > 0) or top-hat (h4 < 0) shape with respect to a Gaussian. Even in spherical potentials, h4 values depend on the orbital anisotropy, the stellar density profile, and the gravitational potential (Gerhard 1993); therefore determining anisotropy and mass distribution from projected velocity dispersion and h4-profiles generally requires dynamical modelling relating all of these variables.

For ETGs with relatively steep stellar density profiles in typical dark matter halos, radial anisotropy in the outer regions leads to radially decreasing projected LOS velocity dispersion profiles and positive h4, caused by an overabundance of stars at zero LOS velocity (Gerhard et al. 1998). This has been analyzed in several lower-mass ETGs (de Lorenzi et al. 2009; Morganti et al. 2013). On the other hand, kinematic studies of the most luminous ETGs ($M_K < -25.7$; Veale et al. 2017) measuring h3 and h4 values out to two effective radii with extended IFS (the MASSIVE survey, Veale et al. 2017) or deep long-slit (Bender et al. 2015) data, found high mean values <h4>≈0.05 at 2 Re while values at the centers are near-zero (e.g., van de Sande et al. 2017). The high h4 values correlate with increasing projected velocity profiles and both are likely caused by mass distributions increasing outwards faster than isothermal (Veale et al. 2018; Wang et al. 2022). We use this to interpret some of the IGL kinematic data below.

In cosmological simulations, the stellar particles found in the simulated galaxies are divided according to two possible origins. They may form out of gas within the host galaxy, either at early times during a period of rapid star formation, or from gas accreted later; in this case they are tagged as the ``*in situ*'' component. Or they are stellar particles formed in satellite galaxies which are later accreted or merged with the host; this second set is labelled ``*ex-situ*''. The accreted stars may be predominantly found at large radii (Oser et al. 2010) but for the most massive galaxies accreted stars dominate at all radii (Pulsoni et al. 2021). The accreted stellar component is characterized by a radial orbit anisotropy (Abadi, Navarro & Steinmetz 2006; Hilz et al. 2012), because the satellites that dissolved and merged with the host come in on predominantly radial orbits. This suggests that the kinematics of simulated galaxies should be more or less radially anisotropic depending on the relative fractions of *in situ* versus accreted (*ex situ*) stars. In Figure 5 we reproduce the anisotropy profiles $\beta(r) \equiv 1 - (\sigma^2_\theta + \sigma^2_\varphi)/2\sigma^2_r$ for the simulated galaxies in Wu et al. (2014), divided into bins of *in situ* fraction. They indeed find that tangentially anisotropic simulated galaxies can only be found in the group with large fraction of *in situ* stars. Almost all simulated galaxies with low *in situ* star fraction (upper panel) are radially anisotropic with $\beta \sim 0$–$0.3$. However, many of the systems with higher *in situ* fraction have similar anisotropies. As a result, Wu et al. (2014) find no correlation between both quantities for their whole galaxy sample. For their radially anisotropic galaxies the $\beta(r)$ profiles are nearly constant with radius, typically with $\beta \sim 0.1$–$0.3$ for R > 2Re. They also find that most slow rotators are radially anisotropic, whereas simulated fast rotator galaxies have both radially and tangentially biased anisotropy profiles. Wang et al. (2022) studied the distribution of $\beta$ values around 1.5 effective radii in ETGs from the TNG simulation (Springel et al. 2018). They find that galaxies with flat or rising outer velocity dispersion profiles are generally radially anisotropic, whereas galaxies with falling outer dispersion profiles can have a wide range of anisotropies at this fiducial radius. These results appear broadly consistent with the findings of Wu et al. (2014).

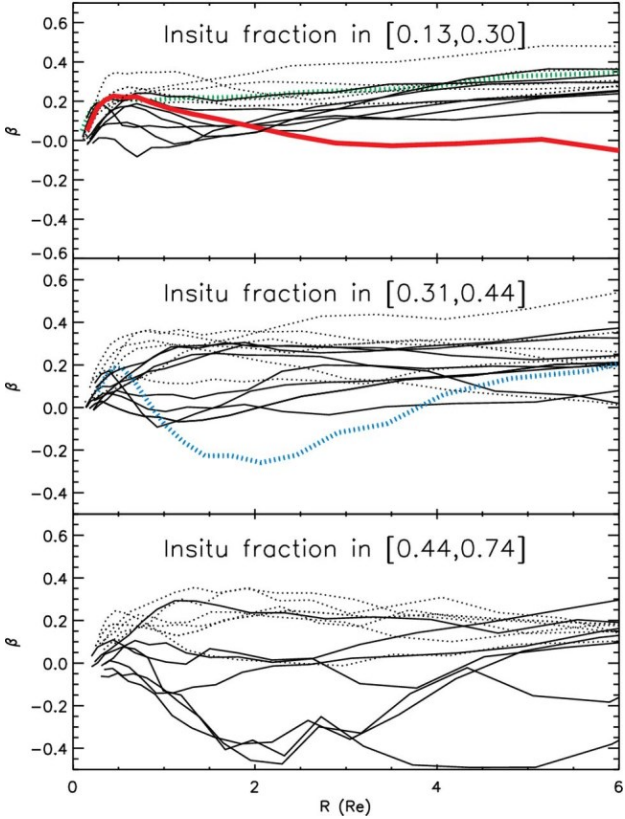

**Figure 5:** Velocity anisotropy profiles for the simulated galaxies from Wu et al. (2014) in bins of *in situ* fraction of stars within 10 per cent $r_{vir}$. Solid lines are for fast rotators while dotted lines are for slow rotators. Colored curves depict the three fiducial galaxies: green dotted line for a slow rotator, blue dotted line for a slow rotator with peaked $\lambda(r)$ profile at 2Re, and red solid line for a fast rotator. Galaxies with low *in situ* fractions have more radially anisotropic orbit distributions, while tangential anisotropy is seen only for systems with high *in situ* fraction. Image reproduced from Wu et al. (2014) with permission.

**5.2 Case studies of nearby groups and clusters**

**Example of increasing circular rotation curve with radius: M49 in Virgo sub-cluster B -** In M49 the rising $\sigma_{LOS}$ profile with radius reaching the velocity dispersion of the Virgo Subcluster B (Binggeli et al. 1993), from 300 - 500 kms$^{-1}$ at 100 – 250 kpc radial distance from the center of M49 arises from the transition between the CCG halo and an outer stellar component with a larger velocity dispersion. These two components overlap along the same LOS following different surface brightness distribution (Spavone et al. 2017). Hartke et al. (2018) constructed the $\sigma_{LOS}$ profile for the faint subsample of PNe in M49, which most likely trace the relaxed smooth component. They used the $v_{LOS}$ of the faint PNe in the radial range 10-80 kpc around M49 to build the global LOSVD; this LOSVD turned out to have strong high-velocity wings. The h4 value is h4 = 0.11 +/-0.03 and a much smaller h3 = 0.01 +/- 0.03 (Hartke et al. 2018). For ordinary massive ETGs, h3 has usually larger values than h4 (Bender et al. 1994). To further illustrate the uncommonly large value of h4 for M49, Hartke et al. (2018) compared the h4 value for the M49 smooth halo to the distribution of luminosity-averaged h4 values of ETGs in the SAURON[4] (Emsellem et al. 2004) and MASSIVE (Veale et al. 2017) surveys. As shown in Figure 6, the large h4 value for the extended PN velocity field in M49 is larger than the h4 for any of these galaxies at smaller radii, including M49 itself (at 2 $R_e$, Veale et al. 2017). The implied positive radial

---

[4] The SAURON survey covers the central 0.5-1 $R_e$, where the h4 values are generally small or even slightly negative (see also Bender et al. 1994).

h4 gradient and the outward dispersion gradient both point to a significant increase of the circular velocity in the transition from the galaxy to the surrounding group.

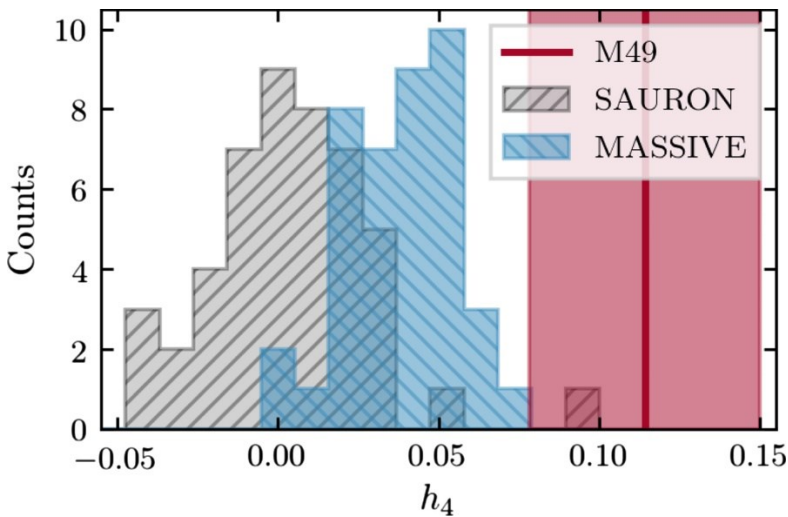

**Figure 6 :** Best-fit $h_4$ value and error band (red vertical line and shaded region) derived from the LOSVD of the M49 faint PN sample compared to the luminosity averaged $h_4$ distribution of 47 and 41 ETGs from the SAURON (Emsellem et al. 2004) and MASSIVE (Veale et al. 2017) surveys (grey and blue histograms). Image reproduced from Hartke et al. (2018) with permission.

We note that the core-winged structure of the LOSVD associated with large positive values of h4 at large radii around M49 can emerge quite naturally from the superposition of two LOSVDs, a narrower and a broader one as predicted by the simulation of CCG and ICL by Dolag et al. (2010).

**Rotation and anisotropy in outer halos –** According to cosmological simulations, both the CCG halos and ICL in clusters result from hierarchical accretion of stars; however, they differ in their spatial distribution, degree of dynamical relaxation and/or phase-space mixing (Cooper et al. 2015). In this section we focus on the outer regions of M87 as an example where ICL and CCG stars can be tagged on the basis of their kinematics at the same spatial location.

The IFS VIRUS-P data of Murphy et al. (2011, 2014) measure a steep increase of $\sigma_{LOS}$ at R≥ 40 kpc in M87. The comparison of these measurements with the $V_{LOS}$ distribution mapped by PNe suggests that the cause for their steep rise in the $\sigma_{LOS}(R)$ profile is the contribution from the ICL. At R≥ 40 kpc, the M87 CCG surface brightness decreases rapidly, while the ICL becomes significant, see Figure 7 which shows the ICL cospatial with the outskirts of M87 and with a large dispersion characteristic of the Virgo cluster core (Longobardi et al. 2015a, 2018a). Once the broad ICL LOSVD and the contribution from the localized substructure (e.g. the crown, Longobardi et al. 2015b) are subtracted, the remaining smooth M87 halo has a LOSVD which is centered on the galaxy's systemic velocity (1307 kms$^{-1}$). The mean rotation velocity of the CCG halo is ~25 kms$^{-1}$, and the velocity dispersion profile, after rising slowly to 300 kms$^{-1}$ at $R_{avg}$ ~50-70 kpc, decreases rapidly to 100 kms$^{-1}$ at $R_{avg}$ ~135 kpc (Longobardi et al. 2018a). In the same region from R>40 kpc the halo surface brightness profile drops below the Sersic fit to a steeper $I(R) \propto R^{-\gamma}$ with $\gamma=(2-2.5)$ and the circular rotation curve obtained from X-ray data (Churazov et al. 2010; Simionescu et al. 2017) turns from flat to steeply rising. Local Jeans models (Churazov et al. 2010; Longobardi et al. 2018a) show that, with this configuration, the measured $\sigma_{LOS}(R)$ profile out to 135 kpc can be reproduced by an isotropic stellar orbital distribution at R≤ 60 kpc which then become *strongly radially anisotropic* in the radial range 70 – 135 kpc.

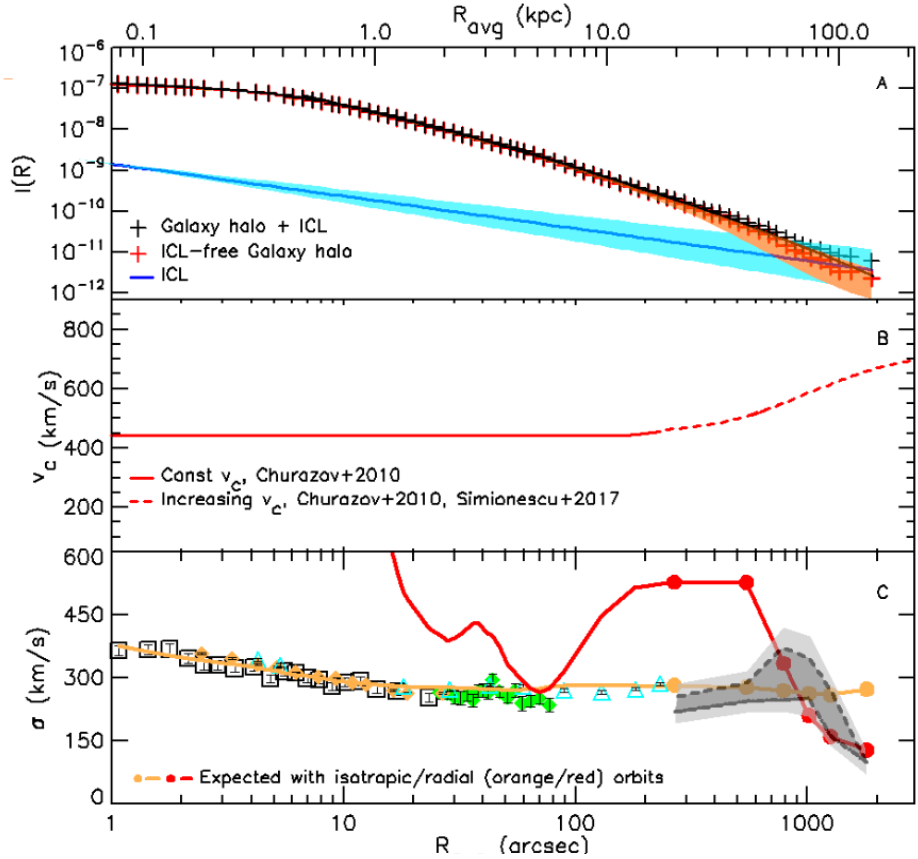

**Figure 7:** Dynamics of the smooth M87 halo out to 2000" compared with predictions from Jeans equations. *Panel A*: surface brightness profile of the M 87 smooth halo (red crosses and shaded area) obtained from the extended photometry (Kormendy et al. 2009), after subtracting the ICL contribution (blue line and shaded area). The black continuous line shows the Sersic fit with $n = 11$ to the total halo+ICL extended photometry. The adopted circular velocity profile based on X-ray data is shown in *panel B*. *Panel C*: velocity dispersion profile $\sigma(R)$ from IFS (squares, triangles) and long slit (diamonds) data (source: Longobardi et al. 2018a). Large dots represent the expected $\sigma$ values at the average radii of the radial bins, for isotropic (orange) and completely radial (red) orbital anisotropy. The comparison with the range of $\sigma$ values for halo PNe (shaded area) suggests that the distribution of orbits changes from near-isotropic at ∼200" to strongly radial at ∼2000". Image reproduced from Longobardi et al. (2018a) with permission.

**NGC 3311 and the Hydra I cluster -** An interesting case where the increasing velocity dispersion profile $\sigma_{LOS}(R)$ correlates with higher moments of the LOSVD is that of the halo in NGC 3311 and the ICL in the Hydra-I core. For NGC 3311, Barbosa et al. (2018) were able to measure the h3 and h4 profiles from the MUSE IFS data, in addition to $V_{LOS}$ and $\sigma_{LOS}$ profiles which are shown in their Figure 16. The skewness h3 profile has about null values out to 2 kpc and then becomes negative at distances larger than 10 kpc, where both $V_{LOS}$ and $\sigma_{LOS}$ increase. The kurtosis h4 profile has also near null values at the center, increases to ≈0.1 at 2-4 kpc, and then decreases to smaller positive values at radii >10 kpc. The positive h4 values measured for NGC 3311 are similar to those measured in other galaxies of the similar luminosity class (Veale et al. 2017). Positive h4 and increasing $\sigma_{LOS}(R)$, as observed in NGC 3311 (see also Hilker et al. 2018) suggest a mass distribution increasingly outwards faster than isothermal ( Section 4.1), and are reminiscent of the properties of simulated massive ETGs in cosmological simulations analyzed by Wu et al. (2014). The most massive galaxies in these simulations have radially increasing circular velocity curves, radial anisotropy in the 2-5 $R_e$ region and a large fraction of accreted *ex-situ* stars (Wu et al. 2014).

**The radial orbital anisotropy of the ICL stars in the Fornax cluster –** In Figure 3, we showed the velocity dispersion $\sigma_{LOS}(R)$ profile of the NGC 1399 halo and the ICL in the Fornax cluster out to 200 kpc. Between 100 and 200 kpc, the $\sigma_{LOS}(R)$ profile is flat at a value $\sigma_{LOS} \approx 300$ kms$^{-1}$, beyond the peak at 400 kms$^{-1}$ reached at ∼ 50 kpc from the center of NGC 1399. When comparing with the velocity dispersion of Fornax cluster galaxies, obtained by averaging in annuli their $V_{LOS}$ at about similar distances from the cluster center, Spiniello et al. (2018) found a negative difference of $\Delta\sigma \sim 80$ kms$^{-1}$ between the velocity dispersion of Fornax satellite galaxies ($\sigma_{cl}$) and that measured for the ICPNe. The lower value for $\sigma_{ICPNe}$ in this radial range is consistent with a dynamical model where the ICPNe orbit in the same potential as the Fornax cluster satellite galaxies, have constant radial anisotropy parameter ($\beta = 0.3$) but have a different, steeper, density profile than that of the Fornax cluster galaxies (averaged in circular annuli). When both distributions are fitted with a power law $R^{-\gamma}$, $\gamma_{ICPNe} = 3.0$ while $\gamma_{cl} = 2.0$ at R=110 kpc, supporting the view that the stars building the ICL are those that were stripped from the satellites on radial orbits that plunge deep into the cluster potential.

# 6 Stellar populations and progenitors of IGL and ICL in nearby groups and clusters

## 6.1 Learning from simulations

Murante et al. (2004, 2007) investigated the spatial distribution and origin of the DSC in simulations in more than hundred galaxy clusters, with masses $M>10^{14}$ $h^{-1}$, which were selected from a cosmological hydrodynamical simulation. Murante et al. (2004) applied a Friend-of-Friend algorithm to remove all satellite galaxies, and then studied the DSC which included CCG and ICL stellar particles. They found evidence for a DSC in the cluster volume once the cluster satellite galaxies (but the CCG) were removed. For each DSC star particle at $z = 0$ in these clusters, Murante et al. (2007) looked for the satellite cluster galaxy to which this particle once belonged at an earlier redshift, thus linking the presence of the DSC stars to the formation history[5] of the cluster. Examples of merger trees are given in Figure 8. The main results of their analysis are that on average, 50% of the DSC star particles come from galaxies associated with the family tree of the most massive galaxy, i.e. the BCG. For the other half, 25% comes from the family trees of other massive galaxies, and the remaining 25% from dissolved galaxies: clearly *the formation of the DSC* is parallel to the build-up of the BCG and other massive galaxies. Most DSC star particles *become unbound during mergers* in the formation history of the BCGs and of other massive galaxies, independent of cluster mass. Murante et al.'s (2007) results further suggest that the *tidal stripping mechanism is responsible only for a minor fraction of the DSC*[6]. At cluster radii larger than 250 $h^{-1}$ kpc, the DSC fraction from the BCG is reduced and the largest contribution comes from the other massive galaxies. In the cluster outskirts, galaxies of all masses contribute to the DSC and most of the star particles come from stripping. An important aspect of ICL formation is the so called preprocessing (Rudick et al. 2006; Contini et al. 2014), by which is meant the formation of diffuse DSC in a smaller entity, e.g. galaxy group, before this later merges with the cluster and releases its DSC to the cluster's ICL when disrupted. For a more extensive discussion of ICL formation processes see the review by Contini (2021).

The DSC does not have a preferred redshift of formation: however, most DSC stars are unbound at $z < 1$ (see also similar results from semi-analytical model predictions by Contini et al. 2014). The amount of DSC stars at $z = 0$ does not correlate strongly with the global dynamical history of clusters and increases weakly with cluster mass.

                                                                                                                                                                        18

---

[5] Galaxies were identified in the simulated clusters at several redshifts, starting with $z = 3.5$, and then the family trees were built for all the $z = 0$ cluster galaxies in Murante et al. (2007). The most massive cluster galaxies show complex family trees, resembling the merger trees of dark matter haloes, while the majority of other cluster galaxies experience only one or two major mergers during their entire life history. Fraction of DSC particles associated with the merger part of the family trees, and with the stripping part of the family trees, were computed as follows. A DSC star particle to arise from a merger at redshift $z_j$ if the galaxy it was last bound to has more than one progenitor at $z_{j-1}$. The DSC particles coming from the progenitors at $z_{j-1}$ are also defined as arising from a merger. A DSC star particle is classified to be unbound through stripping if the galaxy it was last bound to at redshift $z_j$ has only one progenitor at $z_{j-1}$.

[6] The classification of DSC and ICL stars as originating from either ``merger'' or ``stripping'' in simulations depend on the adopted ``operational'' definition. See discussion in the review by Contini (2021).

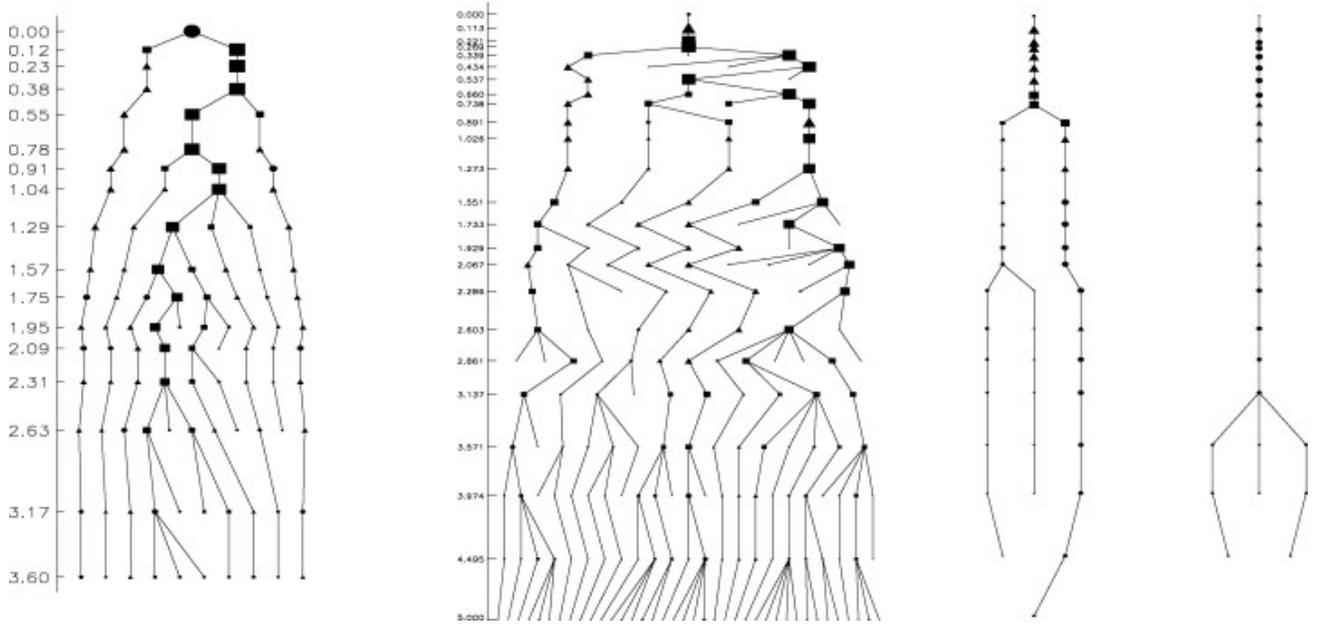

**Figure 8:** Left-hand panel: family tree of the BCG galaxy in a low-resolution simulation. Right-hand panel: family trees of the BCG galaxy, the third most massive galaxy and a lower mass galaxy in the high-resolution re-simulation of the same cluster. The size of symbols is proportional to the logarithm of the mass of the galaxies at the corresponding redshift. Shown on the vertical axis on the left-hand side are the output redshifts used to reconstruct the family trees; these are different in both simulations. A galaxy in these trees is considered a progenitor of another galaxy if at least 50 per cent of its stars are bound to its daughter galaxy, according to the SKID algorithm. Many more galaxies can be identified in the high-resolution simulation at similar redshift. The BCG family tree is characterized by one dominant branch with a number of other branches merging into it, at both resolutions. Squares and triangles are assigned according to the classified 'merging' or 'stripping' events. Circles correspond to redshift at which the galaxy is not releasing stars to the DSC. Image reproduced from Murante et al. (2007) with the permission.

Cosmological simulations like those discussed above show that different galaxies in a cluster have distinct accretion histories with alternative channels contributing to the gravitationally unbound stars which populate the cluster volume. If these different progenitors contribute tracers of different nature, PNe, red or blue GCs, to the DSC with distinct frequencies (which may depend on the stellar masses, hence absolute B band magnitudes of the progenitors and also on galaxy types, see Coccato et al. 2013), the different origin or progenitors may bring about different velocity fields and distinct spatial distributions for the different discrete tracers. Diversity in either the spatial distribution and/or kinematics may provide interesting opportunities to reconstruct the mass assembly, i.e to map which progenitors brought which stars to the DSC and at what radii.

**6.2 PNe as tracers of stellar populations: PN specific frequency and PNLF morphology**

In Section 4 we argued that the CCG halo, IGL and ICL in groups and clusters can be identified according to their different LOSVDs, and thus it is possible to tag tracers according to which kinematic component they belong to at the same spatial position. In this section we discuss the difference in the stellar population properties of the stars kinematically tagged as linked either to the CCG halo or to the IGL and ICL, as from their PN population properties.

In M105, M49 and M87 the kinematically tagged IC and IG PNe show different stellar population properties from the CCG halo PNe (Hartke et al. 2017, 2020; Longobardi et al. 2013, 2015a, 2018a). As a general trend, the inner PN number density profile follows the Sersic fit to the CCG V-band surface brightness profile, while the ICPN number density profile is consistent with a shallower power law, ($N_{ICPN}(R) \propto R^{-\gamma}$) or exponential. In detail, the number density profiles of PNe become ``flatter'' than the surface brightness V band profiles of these bright galaxies in the cluster/group region at radii smaller than those at which the change of slope associated with the excess of light from the IGL or ICL occurs. This distinct outer behavior of IC PNe with respect to the CCG halo population is linked to the different values of the PN specific yields, or α parameter, where $\alpha = N_{tot,PN} / L_{bol}$. The α parameter values of the PN populations associated with CCG, IGL or ICL, are measured empirically and turn out to be different. Current observations show that the IGL and ICL PN populations have several times larger PN yield values per bolometric luminosity than the inner CCG halo light, in the range 3 - 7 time larger. The measured larger values of the PN yield values for the IGL and ICL ($1.6 \cdot 10^{-8}$ - $10^{-7}$ PN $L^{-1}_{BOL}$) are similar to the α-parameter values measured for the PN populations detected in dwarf irregular galaxies of the Local Group, such as Leo I, Sextans A, and Sextans B (Buzzoni et al. 2006). The association of high PN yields with metal poor stellar populations is further explored and supported by the studies of the resolved stellar population in the Leo I group, which is summarized in Section. 6.4.

Another important observational result that supports differences among parent stellar populations is linked to the morphology of the PN luminosity function (PNLF) for these CCG, ICL or IGL PN populations. In the case of M105 and M49, according to the generalized PNLF formula (Longobardi et al. 2015a), the PNLF for the IGL component has a steeper gradient quantified by a larger value of the $c_2$ parameter (in the range 1- 1.5) than that measured for the CCG halo PNLF ($c_2$ value in the range 0.5 – 0.7, Hartke et al. 2017, 2020) or the Ciardullo's original PNLF analytical formula ($c_2$ value equal to 0.302, Ciardullo et al. 1989). The PNLF steeper slope is related with a larger number of core stars with smaller masses, as observed in older stellar populations like those in bulges of nearby galaxies (Ciardullo et al. 2004). In the Virgo cluster, the IC PNLF displays a "dip" (Longobardi et al. 2018a) which is reminiscent of a similar morphological feature detected for the SMC/LMC PNLFs (Jacoby & De Marco 2002; Reid & Parker 2010) and a value of the $c_2$ parameter equal to 0.66.

In summary, the larger PN yield per bolometric luminosity measured for the IGL and ICL and these measured similarities of the IC PNLFs with those observed for low-luminosity, low metallicity, star forming irregulars support the inference that stellar population in the IGL and ICL traced by PNe come from lower mass progenitors than those which deposited the accreted stars at smaller radii within the CCGs.

## 6.3 Resolved stellar population of the ICL in the Virgo cluster

Direct detections of IC red giant branch (RGB) stars in the IC regions around M87 come from the measurement of an excess of unresolved number counts in deep I band (F814W) HST images, with respect to similarly deep cosmological fields (see for example the Hubble UDF). The first case reported of such an excess was measured by Ferguson et al. (1998) in a field 44 arcmin east of M87. Subsequent measurements at independent positions were performed by Durrell et al. (2002) and Williams et al. (2007), in two distinct fields ~40 arcmin north-west of M87. These number count excesses display a sharp edge/increase at the apparent magnitude $m_{F814W} \sim 27$, which corresponds to the apparent magnitude of the tip of the RGBs at a distance modulus (m-M)≈31. From the fit with theoretical stellar evolution models to the color magnitude diagram (CMD) of these sources, it was determined that about 70% of the stars in the WFP2/ACS IC fields are at a common distance of ~16 Mpc (+/-0.1 mag). Adopting a technique which is similar to both Durrell et al. (2002) and Williams et al. (2007), Yan et al. (2008) serendipitously detected an excess of point-like sources in the I band (F814W) HST images available for two ACS fields, at 7 arcmin and 15 arcmin respectively from M60 (NGC 4649). The number density of the excess I band field population shows a decreasing gradient with increasing radial distance from M60. This population of stars around M60 occurs at apparent magnitude $m_{F814W}$ ~ 26.2; its inferred average surface brightness is 26.6 mag arcsec$^{-2}$ in the I band in the field at 7 arcmin from M 60. These values are very similar to those measured by Durrell et al. (2002) observations of the IC RGBs in the Virgo core. Yan et al. (2008) argued that they detected an IG stellar population associated with the M60/M59 galaxy pair. The results linked to the detected excess of I band counts may indicate that the M60/ M59 sub-group is yet another region of the Virgo cluster where IGL is being assembled.

The deep HST/ACS images acquired at the IC positions in the Virgo cluster at 40 arcmin (190 kpc) from M87, and 37 arcmin (170 kpc) from M86 in the wide V (F606W, ~63k secs) and I band (F814W, ~27k secs) filters, were analyzed by Williams et al. (2007). These authors analyzed the I band luminosity function for the ~5300 IC RGB stars (once corrected for background unresolved contaminants) and their CMD distribution. At this sky location, the excess number count population has a corresponding mean V-band surface brightness (excluding background galaxies) of $\mu_V$ ~27.7 mag arcsec$^{-2}$, in agreement with the deep image of the Virgo core from Mihos et al. (2005). The stellar isochrone fits to the CMD showed that the Virgo's IC RGB population is dominated by old, metal-poor stars. The best-fit isochrone models by Williams et al. (2007) indicate that 70%- 80% of the stars have ages that are greater than 10 Gyr; these stars have a median metallicity of [M/H] ~ -1.3 and fall in the range -2.5 < [M/H]< -0.7. However, there is some evidence for the existence of a younger, metal-rich component. Integrated over the entire mass function, the younger component contributes 20%-30% of the total population. The V-I data by Williams et al. (2007) are the deepest CMD available for resolved RGB stars at the Virgo cluster distance to date.

### 6.4 Progenitors of IGL in the Leo I group and Virgo sub-cluster B

The V band profile of M105 in the Leo I group is described by a superposition of a Sersic profile (n=4) and an outer exponential envelope (Watkins et al. 2014; Ragusa et al. 2022). The PNe distribution from Hartke et al. (2020) shows a flattening with radius at large radii, corresponding to an higher PN yield α value in the outer regions. Harris et al. (2007) and Lee & Jang (2016) analyzed HST deep images acquired at two distinct locations, respectively at 4 arcmin and 15 arcmin from the M105 (NGC 3379) center. They assumed a 10 Gyr old age for the resolved stellar population in these fields. This assumption is supported by the general old age of M105 and independently by the steep slope of the PNLFs in the inner and outer regions (Section 6.2). The CMDs show evidence for a metal poor population whose fraction increases outwards, with the number density of metal poor stars with [M/H] < -1.0 exceeding that of the metal richer ones ([M/H]>-1), at approximately ~12 $R_e$ from M105. Using these results, Hartke et al. (2020) constructed a photometric model which conclusively established the association of the exponential envelope with the high PN yield value and the [M/H] < -1.0 metal poor resolved population. In Section 4.2, we already discussed the kinematics association of the exponential envelope with the IGL around M105, which has a σ = 160 kms$^{-1}$. Thus the IGL around M105 must be dominated by metal poor stars with [M/H] < -1.0.

In M49, the CCG of Virgo sub-cluster B, the V band profile (Kormendy et al. 2009; Capaccioli et al. 2015) follows an inner Sersic profile (n=6.0), with an additional flatter, outer distribution. The PN distribution from Hartke et al (2017) shows a flattening at large radii, corresponding to an outer component with high PN yield α as established with a similar photometric model as in M105. In Section 4.2 we presented the evidence that this outer high α component is associated kinematically with the IGL component surrounding M49, with σ = 400 kms$^{-1}$.

In M49, no resolved stellar populations data are available, however Mihos et al. (2013, 2017) published a deep B-V color profile and 2D color map of the halo and IGL out to 100 kpc from the galaxy's center.

The profile shows a strong gradient to bluer color in the outer parts reaching B−V ≈ 0.6 at the largest radius (Mihos et al. 2013). For a 10-Gyr old stellar population, the corresponding metallicity would be [Fe/H] < −1.5, see the simple stellar population tracks in Figure 3 in Mihos et al. (2013). The assumption of an ``old'' age for this component is independently supported by the smooth spatial appearance of the PN (Hartke et al. 2018 ) and light distributions (Capaccioli et al. 2015) which suggest a dynamical age for the IGL significantly older than the orbital precession time, which is at least 5 Gyr around M49 . The blue color, together with the age constraint, and the high α value suggest a similarly metal poor IGL population surrounding M49 as in the Leo I group.

**6.5 The crown in M87: an example of a CCG family tree and extended mass assembly**

A vivid example of an on-going accretion event in a halo of a CCG is the ``crown'' in M87 (Longobardi et al. 2015b). This event is traced by a sub sample of M87 PNe whose projected position-velocity phase space displays an over-density with a distinct chevron-like feature, being the result of the incomplete phase-space mixing of a disrupted galaxy, see Figure 9. For numerical simulations of accretion events with incomplete phase mixing in the M87 halo see Weil et al. (1997). A deep optical image in Longobardi et al. (2015b; processed with un-sharped masking) shows the presence of a crown-like substructure that contributes ~60% of the total light at major axis distances of R ~ 60−90 kpc. The on-going disruption of a recently accreted satellite in the halo of M87 is supported by the azimuthal variation which is observed in the (B-V) color map of M87, with a higher fraction of tracers where the galaxy color is bluer and the velocity substructure has the largest number of tracers. In turn there is a deficit of tracers along the photometric minor axis, where the galaxy is redder, see Figure 2 in Longobardi et al. (2015b). An investigation for kinematic substructures in the projected phase-space of the M87 GCs was also conducted by Romanowsky et al. (2012), which resulted in the discovery of a chevron-like structure also in the projected phase space. Though the morphology in the projected phase spaces for PNe and GCs is alike, they differ in a number of quantitative properties. The width of the chevron for the GCs projected $V_{LOS}$ versus major axis distance (R) phase space goes to zero velocity at $R_{GC}$ ~ 1500″ with $V_{LOS,GC}$ = 1307 kms$^{-1}$, see orange squares in Figure 9, while the width of the chevron for the PNe goes to zero at R ~ 1200″ with $V_{LOS,PN}$ = 1250 ± 21 kms$^{-1}$ , see green circle and magenta diamonds in Figure 9. Furthermore, the GCs associated with the chevron in the GC projected phase space show a different spatial distribution from the selected chevron PNe. The highest density of chevron GCs occur on the NE photometric minor axis of M87 (Romanowsky et al. 2012), while only a few GCs are found near the M87 major axis, where the PNe associated with the ``crown'' substructure are most frequent in number. The identification of a disrupted satellite via the observed ``chevron'' feature in the M87 outer halo's phase-space supports the size growth and late assembly of giant elliptical galaxies' halos.

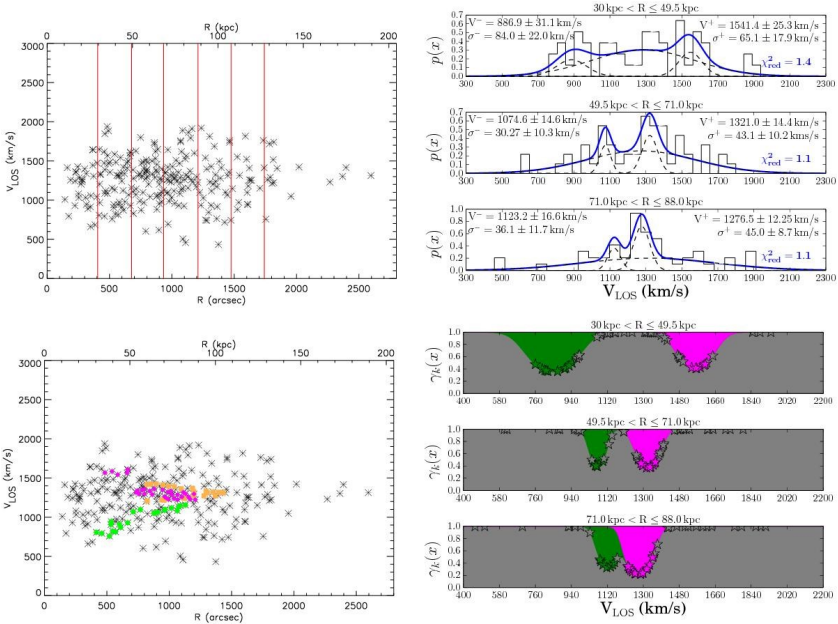

**Figure 9**: *Top left*: projected phase space, $V_{LOS}$ vs. major axis distance ($R$), for all spectroscopically confirmed PNe (black asterisks) in the halo of M 87 from Longobardi et al. (2015a). Red lines separate the elliptical annuli to isolate the cold components associated to the substructure. *Top-right*: LOSVD histograms in the three elliptical annuli. In each panel, the blue lines show the best-fit model computed as a combination of three Gaussians. Black dashed lines show the relative contribution of each component to the LOSVD, with parameters for the cold components given in the plot. *Bottom left*: as in the *top left panel*, however, the green circles and magenta diamonds show the PNe associated with the cold secondary peaks in the LOSVD. Orange squares show a kinematically selected GC substructure from Romanowsky et al. (2012). *Bottom right*: probability that a PN is drawn from the halo component (dark grey area) or from the chevron (green, magenta areas). Stars represent PN probabilities at their measured $V_{LOS}$. Image reproduced from Longobardi et al. (2015b) with permission.

## 6.6 Stellar populations of the ICL around M87 in the Virgo cluster core

In M87, the CCG of Virgo sub-cluster A, the V band profile (Kormendy et al. 2009) follows a Sersic profile with high index (n=10.0), out to 150 kpc major axis distance, with the two furthest photometric points reflecting the additional contribution from the ICL (Mihos et al. 2005, 2017). Longobardi et al. (2013) showed that PN distribution can be decomposed into an inner halo component following the Sersic profile and an outer ICL component with 3 times larger PN yield α. From the kinematic decomposition described in Section 4.2, Longobardi et al. (2018a) showed that the number density distribution of the PNe identified as the ``smooth'' M87 halo follow the same Sersic profile as the light, and the ICPN component, which includes the unrelaxed component of the M87 halo and the broad σ=900 kms$^{-1}$ component, follows a power law R$^{-γ}$ with γ = 0.79 ± 0.15.

A 2D color map for M87 and its immediate surroundings was published by Mihos et al. (2017) which showed a strong color gradient in the outer M87 halo, reaching colors consistent with B-V= 0.6 at 135 kpc major axis distance. The higher PN yield α ICL component dominates the surface brightness at these distances around M87. We know from the previous IGL discussion in Sect 6.4 that this is associated with metal poor stellar population. This inference is independently supported by the best-fit isochrone models by Williams et al. (2007) to the ACS field CMD between M87 and M86. These indicate that 70%- 80% of the stars in this field have ages greater than 10 Gyr and a median metallicity of [M/H] ~ -1.3, in the range -2.5 < [M/H]< -0.7. In addition Williams et al. (2007) found a 20% contribution from a population with [M/H] ~ -0.3 and younger ages (2-3 Gyr). The HST-ACS data were later re-analyzed by Lee & Jang (2016) assuming a 10 Gyr age, α-enhanced ([α/Fe] =0.2) population using Dartmouth isochrones (Dotter et al. 2008). In their analysis they found a larger contribution from subsolar metallicity stars, about 50%, similar to their results for the outermost field from M105 in the Leo I group.

In the case of the M87 halo and surrounding ICL, the dynamical age is not constrained. However the PNLF of the IC PNe has steep slope consistent with an older age but at the same time shows the ``dip'' observed for star forming populations making a case for a contribution from stars with younger ages in this area. Thus the combined properties of the Virgo ICPN population near M87 appear to be most readily explained if this population derives from a faded population of low-luminosity, low-metallicity, star-forming or irregular galaxies, such as M 33 or the LMC, which are very different from M 87 itself, possibly with additional stars from more massive galaxies that later merged with M87. Longobardi et al. (2018a) found that the IC stars sampled in their survey fields correspond to ∼1.5 M 33- or ∼4 LMC-like galaxies.

### 6.7 Stellar population constraints on ICL in massive clusters from absorption line spectra

Within the limit imposed by long integration times (to reach suitable S/N) because of the steeply decreasing surface brightness profiles, efforts are carried out to constrain the metal content of stars in the CCG halos and ICL, using stellar absorption line features of the stellar continuum. In general stars in massive ETGs at 2 $R_e$ or further out have high [α/Fe] (Green et al. 2013; Bender et al. 2015) and produce measured profiles with declining metallicity gradients (Zibetti et al. 2020) and flat-to-positive alpha element gradients (Kuntschner et al. 2010; Coccato et al. 2010, 2011). The high [α/Fe] is associated with stellar populations formed during a short, intense burst of star formation (Thomas et al. 2005). Negative metallicity gradients which extend to several $R_e$ are found in most ETGs (e.g. Baes et al. 2007; Spolaor et al. 2010; La Barbera et al. 2012).

As an example, the outer envelope of NGC 3311 provides evidence of un-mixed stellar populations with different [α/Fe] values and metallicity (Barbosa et al. 2016, 2021). At the radii where the σ increases, the stars in the off-centered exponential envelope are younger that those in the central regions, with lower metallicity but α-enhanced. Barbosa et al. (2021) stated the because of the central location of NGC 3311 in the Hydra I cluster the accreted stars which build the ICL come mostly from rapidly quenched satellites.

## 7 Concluding remarks and forward look

**What have we learnt on ICL and IGL from studies in the local groups and clusters?** The study of the ICL and IGL in the local groups and clusters in the nearby Universe allows us a privileged view on these components and their phase space properties, which is complementary to the photometric studies carried out for clusters at higher redshifts. In the nearby Universe it is possible to measure the projected phase space – radius and velocity – for a sparse but still significant sample of individual IGL and ICL tracers. Nearby clusters are also within reach of HST, and pointed observations provide specific information on the age and metallicity distribution function of the resolved stellar populations in these clusters and groups.

The IGL and ICL surface brightness distributions in the nearby clusters are more centrally concentrated than the azimuthally averaged light of group and cluster galaxies excluding the CCG: by itself it is an indication that stars are gravitationally unbound and added to the group or cluster volume from those galaxies that plunge deeper into the group or cluster potential. The ICL and IGL fractions are a few to ten percent of the light in cluster and group galaxies. These fractions are lower than those measured in more concentrated, dynamically evolved clusters at larger distances. The ICL stars are not bound to a single galaxy, but by the cluster or group potential. The binding energy of the ICL may be in fact be a continuum depending on the whether the cluster is relaxed/compact or affected by sub-cluster mergers. From the measured CMD in deep HST images, the theoretical stellar evolution models suggest that the IC population is dominated by stars with old ages (≥10 Gyr) with a significant number of very metal-poor stars ([Fe/H] < −1.5). See also the very recent results on the color gradients in the ICL towards bluer color in intermediate redshift (0.2-0.8) clusters (Golden-Marx et al. 2022).

Both the observations and the simulations show that the ICL *in situ* star formation from stripped gas (Puchwein et al. 2010) is not an important channel, see discussion in Contini (2021). But it does happen: see the tidally stripped gas in the Virgo core (Oosterloo & van Gorkom 2005) and the isolated HII around NGC 4388 (Gerhard et al. 2002). Hence the majority of IC stars in the nearby groups and clusters were formed in galaxies and then unbound in the cluster space.

**What are the likely masses of the satellite progenitors of the ICL and IGL stars?** From the arguments presented in Section 6, the picture that emerges for the local IGL/and ICL components is that their stars are mostly old and metal poor implying low mass satellite progenitors. The stellar masses of the satellites that built the IGL component around M105 or M49 would have to be a few $10^8$ M$_\odot$, according to the low-mass end of the mass-metallicity relation (Kirby et al. 2013). In more massive clusters, we learn from the NGC 3311 case in the Hydra I cluster that a larger percentage of stars will come from more massive, metal richer progenitors with faster star formation time scale (high [α/Fe]).

An evolutionary path for the formation of a blue, smooth, and extended IGL around M49/M105/M87 may start with an early (before z ∼ 0.5) accretion of a number of low-mass satellites with young (∼1 Gyr) and blue (B − V = 0.1) stellar populations at that redshift. Since their accretion, more than 5 Gyr ago, these stars would have evolved passively and now reached a color of B − V = 0.65 at z = 0. The presence of stars that originate from low-mass systems is related to the presence of large amounts of dark matter in groups and clusters. Due to dynamical friction, low-mass systems will stay roughly at the radius at which they were disrupted, while the more massive and therefore metal-rich/redder satellites would sink towards the CCG center (e.g. Ostriker & Tremaine 1975; Amorisco 2017). We note that the contribution of such low-mass systems with small mass ratios with the host galaxy to the ICL and IGL is expected to be small in current hydrodynamical cosmological simulations. While several groups/authors are able to carry out simulations with mass resolution equivalent to mass merger ratios $10^{-4}$ : 1, the cumulative effect from mergers with stellar mass ratios below $10^{-2}$ : 1 is found to be negligible with respect to stellar halo growth (see Fig. 1 in Rodriguez-Gomez et al. 2016). The observations of gravitationally unbound IGL and ICL in the nearby groups and clusters suggest instead that the IGL and ICL formed early from the accretion of many low-mass satellites. Therefore this channel which adds old metal poor stars to the IGL and ICL is implied in the nearby universe.

**Observations of higher redshift clusters –** The results illustrated in the previous sections on the origin of stars in the ICL refer to those stars identified because of their i) large relative velocities which differentiate them from CCG halo stars and ii) distinct properties in the CMD. In clusters at higher redshifts/distances, measurements of the detailed kinematics and stellar population properties are challenging, hence the photometric approach is the one most often implemented observationally. For observations and characterization of the DSC in higher redshift clusters see Montes et al. (2021), De Maio et al. (2015, 2020) and references therein. Hence because of observational limitations, only comparatively higher surface brightness regions of the DSC are accessible in these clusters. These are then related to the stars stripped from more massive companions or liberated during massive merger events (Purcell et al. 2007; Contini et al. 2014, 2017).

**Future outlook -** As photometry becomes deeper with the LSST, or carried out in space with Euclid, it may be possible to detect the outer boundary of the ICL, i.e. the splashback/ truncation radius (Gonzalez et al 2021.) and thereby constrain the redshift of the earliest infall. An additional strong motivation to map and trace the DSC, i.e. CCG halo light, IGL and ICL, comes from the observational results showing that the (dark) matter distribution from weak lensing and the DSC from deep photometry map each other (Montes & Trujillo 2019) and the results from cosmological simulations indicating that the kinematic properties of stars and dark matter in ETG halos become similar at large radii where the accretion factions are largest (Pulsoni et al. 2021).

The new facilities ELT/JWST/Euclid will revolutionize the study of the IGL and ICL. With integral field spectroscopy on the ELT it will be possible to constrain the ages, metallicity and [α/Fe] in the Virgo core. An opportunity may come directly from the nebular phases of the stellar evolution, PNe and HII regions: from their oxygen and argon abundances (as in Arnaboldi et al. 2022 and Bhattacharya et al. 2022), one will be able to constrain the star formation efficiencies of the progenitors that released stars to the ICL. With ELT and JWST it will be possible to study further the resolved stellar populations in clusters and groups, to constrain the CMDs of the ICL stars in different regions of the nearby clusters, thus leading to a better understanding of this enigmatic component.

## 8. Acknowledgments

The quest for discrete tracers in the nearby clusters have benefitted from the contribution and discussion with several colleagues. Special thanks to Ken C. Freeman, Sadanori Okamura, and Matthew Colless for advice and support at the earlier stages of this project. Many Ph.D students and Post docs (now colleagues) contributed to these investigations: J.A.L. Aguerri, C. Barbosa, S. Bhattacharya, N. Castro-Rodriguez, L. Coccato, A. Cortesi, J. Hartke, A. Longobardi, N.R. Napolitano, M. Pannella, C. Pulsoni, C. Spiniello, G. Ventimiglia. We acknowledge collaboration with A. Buzzoni, M. Capaccioli, R. Ciardullo, N. Douglas, J. Feldmeier, G. Jacoby, K. Kujken, R. Kudritzki, R. H. Méndez, M. Merrifield, J.C. Mihos, G.M. Murante, A. Romanowsky, M. Roth, N. Yasuda. Finally we thank the referees and the editor, E. Contini, for comments and suggestions that improved the content of this review.